**Community College Articulation Agreement Websites: Students' Suggestions for New Academic**

**Advising Software Features**

David Van Nguyen[1], Shayan Doroudi[1, 2], and Daniel A. Epstein[1]

[1] School of Information and Computer Sciences, University of California, Irvine, USA

[2] School of Education, University of California, Irvine, USA

**Author Note**

We have no conflicts of interest to disclose. Regarding datasets, this paper and our other related paper use data from the same dataset. However, each paper presents distinct results. This paper uses the dataset's non-experimental data on ASSIST software feature suggestions. Whereas our other related paper uses the dataset's experimental data on algorithm-generated versus human-generated academic plans (D. V. Nguyen et al., 2023).

**David Van Nguyen:** Conceptualization, Funding Acquisition, Methodology (Lead), Investigation, Formal Analysis, Writing - Original Draft, Writing - Review & Editing (Lead). **Shayan Doroudi:** Supervision (Equal), Methodology (Supporting), Writing - Review & Editing (Supporting). **Daniel A. Epstein:** Supervision (Equal), Methodology (Supporting), Writing - Review & Editing (Supporting).

Correspondence should be addressed to David Nguyen. Email: dvnguye5@uci.edu

**Abstract**

Articulation agreements provide more transparency about how community college courses will transfer and fulfill university requirements. However, the literature displays conflicting results on whether articulation agreements improve transfer-related outcomes; perhaps one contributor to these conflicting research results is the subpar user experience of articulation agreement reports and the websites that host them. Accordingly, we surveyed and interviewed California community college transfer students to gather their suggestions for new academic-advising-related software features for the ASSIST website. ASSIST is California's official centralized repository of articulation agreement reports between public California community colleges and universities. We analyzed the open-ended survey and interview data using structural coding and thematic analysis. We identified four themes around students' software feature suggestions for ASSIST: (a) features that automate laborious academic advising tasks, (b) features to reduce ambiguity with articulation agreements, (c) features to mitigate mistakes in term-by-term course planning, and (d) features to facilitate online advising from advisors and student peers.

**Community College Articulation Agreement Websites: Students' Suggestions for New**

**Academic Advising Software Features**

The community college transfer process is inundated by issues that impact transfer rates (Jenkins & Fink, 2016), time-to-transfer (Packard et al., 2012), excess course credits (Fink et al., 2018; Zeidenberg, 2015), transfer credit loss (Giani, 2019), transfer credit applicability (Hodara et al., 2017), and time-to-bachelor's degree  (Lichtenberger & Dietrich, 2017). A policy solution that may directly or indirectly address these aforementioned transfer issues is articulation agreements, which are "formal arrangements that establish course equivalencies and the transferability of academic credit in an effort to facilitate seamless transfer of students' credit across postsecondary institutions" (Crisp, 2021, p. 61).

At universities without articulation agreements, university staff (i.e., transfer credit evaluators) will use an applicant's transcripts and/or course syllabi to determine if the university will accept the community college course and if accepted whether that course credit will fulfill major, general education, or elective degree requirements (Ott & Cooper, 2014). Without formalized articulation agreements, community college stigma can potentially influence these ad-hoc transfer credit evaluations (i.e., community-college-to-university transfer students' courses may be judged more stringently than university-to-university transfer students' courses) (Bowker, 2021; Hyatt & Smith, 2020; D. V. Nguyen, 2023). In many cases, universities will only perform these evaluations *after* applicants are accepted into the university and have submitted their intent to enroll deposit; this timing does not allow applicants to make informed college choice decisions (Grote et al., 2019; Ott & Cooper, 2014).

Although articulation agreements are an improvement compared to scenarios without articulation agreements, the literature displays conflicting results on whether articulation agreements improve transfer-related outcomes (Anderson et al., 2006; Boatman & Soliz, 2018; LaSota & Zumeta, 2016; Roksa & Keith, 2008; Schudde et al., 2023; Spencer, 2019; Stern, 2016; Worsham, DeSantis, et al., 2021; Worsham, 2022). Perhaps one contributor to these mixed results is the subpar user experience of articulation agreement reports and the websites that host them. For example, past research suggests that articulation agreements are difficult to locate (U.S. Government Accountability Office, 2017), require reading levels that are too advanced (Taylor, 2017), contain undefined administrative jargon (Reeping & Knight, 2021), and are confusing (Logue et al., 2023).

For this study, we surveyed and interviewed California community college transfer students to gather their suggestions for new academic-advising-related software features for the Articulation System Stimulating Interinstitutional Student Transfer (ASSIST) website. ASSIST is "the official statewide database [of articulation agreements] that shows … [how California] community college [courses] may be used to satisfy elective, general education and major requirements" at California State University (CSU) and University of California (UC) campuses (ASSIST, n.d., para. 2).

## Related Work

In the related work section, we discuss relevant literature on (a) user experience of articulation agreement reports and websites and (b) education technology prototypes related to academic advising.

**User Experience of Articulation Agreement Reports and Websites**

Among other sources, students and academic advisors[1] rely heavily on websites for transfer-related information (Grote et al., 2019; Holliday-Millard, 2021; Schudde et al., 2021). For example, during appointments, advisors help community college students navigate university websites and go over online articulation agreements (Holliday-Millard, 2021). Yet, many 2-year and 4-year colleges websites have transfer-related information that is missing, incomplete, outdated, confusing, and/or difficult to locate (Bailey et al., 2017; Fincher et al., 2014; Holliday-Millard, 2021; Logue et al., 2023; Schudde et al., 2020; U.S. Government Accountability Office, 2017).

In contrast to subpar websites, a website analysis of community colleges – that were nationally recognized for their work in promoting student success – found that transfer information was accessible by being placed prominently on or near the homepage on their websites. This information was also located in various locations on the websites, implying its significance … The content was presented on each site using various forms of media such as audio, visual and interactive tools. Each website was comparably organized with information on major program maps, a connection to transfer resources, and specifically individuals who can help. (Patel & Brosnan, 2022, p. 101)

Research suggests that the majority of articulation agreements are written at a post-bachelor's degree reading comprehension level, which exceeds the reading level of community college students (Taylor, 2017). Furthermore, research suggest that articulation agreements commonly use (a) undefined administrative jargon which reduces understandability for student readers and (b) "language to instill doubt about the transferability or applicability of student credits" (Reeping & Knight, 2021, p. 332).

---

[1] The terms counselor and advisor are often used interchangeably but may mean different things depending on the college. For clarity, we will use the term advisor: "Academic advisors assist with goal setting, degree planning, and major and career exploration" (Menke et al., 2020, p. 85).

Studies suggest that articulation agreement reports often have inconsistent formatting between universities, which can make it difficult to efficiently cross-reference multiple reports to develop an academic plan (Holliday-Millard, 2021; Taggart et al., 2000). Moreover, users (a) perceived that articulation agreements have confusing formatting and (b) recommended that information within articulation agreements be presented more succinctly (Holliday-Millard, 2021; Taggart et al., 2000). Lastly, Excel file formats were difficult to read and users instead preferred PDF file formats (Holliday-Millard, 2021).

We now discuss *fragmentation* of articulation agreements within a state's public higher education system: "Fragmentation captures the degree to which the information needed to make an optimal decision or action is localized—that is, if the information is all in one place or is 'fragmented' into pieces across several [web]pages" (Reeping & Knight, 2021, p. 319). Users have a relatively easy time locating articulation agreements when all the state's articulation agreements are located within a single centralized website (Taggart et al., 2000). For example, all CSU and UC campuses have their articulation agreements located in the ASSIST website (Taggart et al., 2000). On the other hand, users can have a difficult time locating articulation agreements within their states' public higher education system when they are scattered across each university's institutional website (Holliday-Millard, 2021; Katsinas et al., 2016). For example, major articulation agreements for University of North Alabama and University of South Alabama are located in different websites (Katsinas et al., 2016).

**Academic-Advising-Related Prototypes**

Shifting from academic advising problems to potential solutions, researchers have developed various prototypes that incorporate technology into academic advising. However, few prototypes focus on community college academic advising. The following prototypes have been developed for non-transfer university students. Several prototypes focused on offering recommendations: recommend majors (Obeid et al., 2018), recommend courses (Chang et al., 2022; Ganeshan & Li, 2015; Pardos & Jiang, 2020), and recommend optimal academic plans (Mohamed, 2016; Shakhsi-Niaei & Abuei-Mehrizi, 2020; Xu et al., 2016). Other prototypes focused on predicting workload (Borchers & Pardos, 2023) and predicting grades (Castells et al., 2020; Mendez et al., 2021) based on a student's intended schedule of courses. Prototypes developed to improve advisor-specific tasks include identifying students at risk (Fiorini et al., 2018; Lonn et al., 2012), deciding the upcoming semester's course offerings (F. Lin et al., 2012; Nachouki & Naaj, 2019; Robledo et al., 2013), extracting insights from advisor's written notes (Yang et al., 2021), and offloading academic advising to chatbots (Kuhail et al., 2023; Mekni et al., 2020; H. Nguyen et al., 2023).

We now shift to research prototypes specifically focused on articulation agreements; these agreements are vital for community college students' academic planning. Pardos et al. (2019) and Heppner et al. (2019) developed similar prototypes that use semi-automation to achieve the same goal: reduce articulation officer's[2] workload with the time-intensive process of developing course articulations. Both prototypes output reports that suggest potential course articulations between two colleges. For each course articulation suggestion within the report, articulation officers can either immediately disregard obvious false positives or devote human labor to further investigate the plausible suggestions.

---

[2] To explain their role, articulation officers are "responsible for receiving articulation requests, choosing which to consider, and then beginning the process of validation by conferring with the instructor of record at the other institution by way of its respective articulation officer" (Pardos et al., 2019, p. 1).

## Methods

We conducted a qualitative study to gather California community college transfer students' suggestions for new academic-advising-related software features for the ASSIST website. Our qualitative data consisted of open-ended survey data and semi-structured interviews.

Before conducting the study, we pilot tested the study's survey and interview with three transfer students. We then revised the study based on the pilot test feedback.

### Participants

We recruited 24 community college transfer students through the CSU and UC campuses' subreddit, which are online forums. The main participant eligibility requirements were (a) at least 18 years old and (b) current community college transfer student at a UC or CSU campus. The participant incentive was a $15 gift card.

Table 1 contains the participant demographics. Furthermore, when provided with 5-point unipolar response options, students on average rated their prior knowledge of ASSIST as "very knowledgeable" ($M = 4.0$, $SD = 0.7$). During the results section of this paper, participants will be referred to as P1 – P24.

**Table 1**

*Participant Demographics*

| Demographic | *n* | **%** |
|---|---|---|
| Gender | | |
| Female | 13 | 54 |
| Male | 11 | 46 |
| Race and ethnicity | | |
| Asian | 10 | 42 |
| Hispanic or Latino | 5 | 21 |
| Middle Eastern or North African | 1 | 4 |
| Native Hawaiian or Other Pacific Islander | 2 | 8 |
| White | 10 | 42 |
| Age | | |
| 19-22 years old | 20 | 83 |
| 23-26 years old | 2 | 8 |
| 27-30 years old | 2 | 8 |
| Years enrolled in community college | | |
| 1-2 years | 15 | 62 |
| 3-4 years | 8 | 33 |
| 5-6 years | 1 | 4 |

**Procedure**

Research sessions were conducted through Zoom videoconference between November 2021 to December 2021. The scheduled length of the research session was 45 minutes. First, study participants completed an experimental task where they created an optimal academic plan, which is outside the scope of this paper but is discussed in our other related paper (D. V. Nguyen et al., 2023).

Then, participants completed the survey, which was administered via Qualtrics. Aside from demographic questions, there were three open-ended survey questions that are relevant to this specific paper. Two questions asked for suggestions to improve ASSIST and one question asked for concerns they had with ASSIST.

Finally, if the research session had time remaining, we then conducted semi-structured interviews with study participants to gather their software feature suggestions for ASSIST. 22 out of 24 study participants were interviewed. We told participants not to worry about real-world constraints such as feasibility or cost when making software feature suggestions.

For each participant quote in the results section, we will designate whether the researcher or the participant initiated the conversation about the software feature topic. If the participant quote was researcher initiated, we will state the initiating interview question in a footnote.

When participants initiate the conversation of a particular software feature, we may have asked follow-up questions to get participants to elaborate. Under that circumstance, participant responses to follow-up questions will still be marked as participant initiated because the participant was the one who first broached the software feature topic.

**Analysis**

We used Otter.ai to obtain automated transcriptions of our interview audio. We then manually reviewed the Otter.ai transcripts to fix automated transcription mistakes. We also took a *denaturalized* approach to our transcripts (Oliver et al., 2005). In other words, we removed false starts, repeated words or phrases, and filler words in order to improve clarity without changing the speaker's original meaning.

We analyzed our qualitative data using the ATLAS.ti software. We specifically used *structural coding* to code the open-ended survey data and interview data (Saldaña, 2013). We then conducted *thematic analysis* on the codes (Braun & Clarke, 2006). In other words, we coded interview and survey responses based on the software feature they represent. Then we categorized software feature codes into conceptually similar higher-level themes.

**Positionality**

We limit this brief discussion of positionality (Holmes, 2020) to just the lead author as he was the one responsible for collecting and analyzing the qualitative data. The lead author did not have a pre-existing relationship with any study participants. The lead author was a California community college alumnus who has personal experience using ASSIST to prepare to transfer to UC and CSU campuses. To be transparent, the lead author had a preconception going into the study that the ASSIST website provides a poor user experience and needs improvement. Among other sources, he drew from his personal experiences when developing the interview guide. Furthermore, his positionality likely influenced how interviews were conducted and analyzed (e.g., an interviewer without personal experience with ASSIST may have asked more follow-up questions to clarify when interviewees use ASSIST-related terminology or talk about common ASSIST use cases).

**Limitations**

One of the primary limitations is that our qualitative dataset is relatively small. In general, students only wrote one or two sentences for the open-ended survey questions. Amongst all participants interviewed, we only collected 6 hours and 20 minutes of total interview data. The average length of an interview was about 17 minutes. In retrospect, we should have made the overall research session longer to allow for additional time for the research session's interview portion.

Second, the transfer students who participated in our study may not be generalizable to California community college students. For example, 62% of our study participants transferred within 2 years. For context, among California community college students who aspire to transfer, only 5% transfer within 2 years (Johnson & Mejia, 2020). Moreover, White and Asian students are overrepresented among our study participants, whereas other racial/ethnic groups such as African American or Hispanic students are underrepresented (California Community Colleges Chancellor's Office, 2022). A lack of input from underrepresented minorities is problematic considering that research suggests racial disparities exist in (a) access to high-quality academic advising (Ocean et al., 2014; Orozco et al., 2010), (b) the policy impacts of articulation agreements (Spencer et al., 2023; Worsham, Whatley, et al., 2021), and (c) transfer-related outcomes (Crisp et al., 2020; Y. Lin et al., 2023).

Third, study participants who majored in computer science (8 out of 24 participants) are overrepresented in terms of the proportion of quotes included in the manuscript. As a broad generalization, the computer science students in our study tended to have an easier time describing software features and their quotes were more eloquent.

## Results

We identified four themes around students' software feature suggestions for ASSIST: (a) features that automate laborious academic advising tasks, (b) features to reduce ambiguity with articulation agreements, (c) features to mitigate mistakes in term-by-term course planning, and (d) features to facilitate online advising from advisors and student peers.

### Features that Automate Laborious Academic Advising Tasks

When done manually, certain articulation-agreement-related tasks are too time-consuming that it (a) may not be possible for a community college advisor to complete within a 30-minute advising appointment or (b) may deter students from performing such tasks on their own. This theme – features that automate laborious academic advising tasks – has two software feature suggestions: (a) query other community colleges that offer the university course equivalent and (b) query university major programs that the student is close to completing.

#### *Query Other Community Colleges that Offer the University Course Equivalent*

A common articulation agreement pain point is that "the [student's community college] doesn't have … courses that are equivalent" (P20, participant initiated) to a specific university major requirement, which the ASSIST report designates as "No Course Articulated" (P19, participant initiated). In other words, these students were unable to fulfill "all of the lower division [university major] requirements at [their] specific [community] college" (P6, participant initiated).

Accordingly, students have the option of fulfilling university major requirements at community colleges other than their primary community college. However, P3 (participant initiated) said "it's not really informed to us that some [community colleges] have more classes available for … transfer than others. … That's not really well known." P19 (participant initiated) illustrates the prevalence of this ASSIST use case, presumably by students who were savvier with ASSIST: "There wasn't a lot of STEM classes that were articulated [at my community college] … And so, a lot of people would have to take them at different community colleges that were nearby."

P20 (participant initiated) enrolled at a different community college to take a course not available at his primary community college in order to remain competitive with other transfer applicants: "I would actually feel worried [if I did not fulfill the major requirement] because I felt like I will be behind the other [transfer applicants who have fulfilled the major requirement]." However, some students still enroll in other community colleges even if their primary community college does have the university course equivalent. For example, the primary community college's "class is full or it's not offered in the right time slot" (P9, participant initiated).

Currently, searching for a specific articulated course at other community colleges is a manual process. P21 (researcher initiated[3]) said: "[I create a list of] the community colleges in my area. And [then I] manually [retrieve] the agreements for each [community college in my list]. And then [I see] if any of them have [the specific] articulated course." Once the student finds the specific articulated course at another community college, they then need to check (a) if the course will be offered in the upcoming term and (b) if they are eligible to enroll: "Then you had to manually go to that community college's website. See if they're offering [the articulated course] next semester. … [Then check] can you actually register … like if there's prerequisites

---

[3] For participants who were prompted, the interview question that initiated the conversation around this software feature was along the lines of "Sometimes there are no equivalent courses at your community college to satisfy a university major requirement. However, there may be an equivalent course at another community college. So, what ASSIST functionality would be helpful for this situation?"

on that class" (P23, researcher initiated). P23 (researcher initiated) said this total process took

him "four or five hours, at least."

Students suggested an automated feature where users can input a "[university] course that they need [community college course] equivalents for" and then ASSIST would output "all the courses … from different [community colleges] that [articulate] to that [university] course" (P20, participant initiated). Furthermore, students suggested an option to "filter [equivalent community college courses by] if it's online, in person, hybrid" (P23, researcher initiated). Furthermore, if the course has an in-person component then there should be an option to "sort [equivalent community college courses] by area" (P9, participant initiated) or distance "because you don't want it to be like a three-hour commute" (P12, researcher initiated).

### Query University Major Programs that the Student is Close to Completing

In order to speed up time-to-transfer, students suggested a feature to query university major programs that the student is close to completing. Assuming that the student is only interested in one specific major (e.g., biology) and not picky with the university they transfer to,

then one potential use case is to sort all UC and CSU campuses by the percentage of completed

major requirements for that one specific major. P5 (participant initiated) uses a student majoring

in history as an example:

> If they were taking a bunch of history classes … maybe ASSIST can generate …
>
> [university history] programs they can look into. … [The history programs can be sorted
>
> by] the shortest amount of [uncompleted lower-division major] requirements a school
>
> has.

In another use case, a student could query all the UC and CSU campuses for major requirement completion percentages for all possible majors or restrict it to a user-determined set of similar majors. For example, computer science, computer engineering, software engineering, electrical engineering, and data science are relatively similar. P23 (participant initiated) said:

> If you're [majoring in] CS, you could potentially go to [major in] math [instead], because there's a lot of overlap [in lower-division major requirements]. You could potentially have a feature where [ASSIST outputs] like, "You've completed 50% of the major requirements for [UC Riverside] major B. But you've also completed 75% of [UC Riverside] major C."

P23 (participant initiated) states "[This feature] might actually convince people to just pursue a different major if they see they're very close to finishing [a different but related] major as opposed to their intended major." For example, P13 (participant initiated) was initially a history major but decided to switch to sociology (a relatively similar major) because it had fewer major requirements: "When I first started community college, I was considering being a history major … I [initially] picked like four CSU colleges to apply to and it was so many [history major] classes I have to complete before I transfer."

## Features to Reduce Ambiguity with Articulation Agreements

This theme – features to reduce ambiguity with articulation agreements  – has four software feature suggestions: (a) consistently label major requirements as mandatory or recommended, (b) consistently publish annual ASSIST reports, (c) notification system for ASSIST report changes, and (d) track progress towards completing university transfer requirements.

### *Consistently Label Major Requirements as Mandatory or Recommended*

For context, major requirements on ASSIST reports can be mandatory or recommended. "Mandatory courses … has to be taken … before transferring" (P20, participant initiated). On the other hand, "highly recommended [major requirements] just means … you could finish it up now [at community college], or you could ... [finish it later after] you enter the university" (P18, participant initiated). While not mandated, students feel motivated to complete recommended major requirements during community college to become "a top applicant to get into the [university]" (P20, participant initiated).

ASSIST does not consistently label which major requirements are mandatory or recommended. For example, P16 (researcher initiated[4]) said "I found it kind of difficult to tell which ones were required versus which ones were recommended." P19 (researcher initiated) explains how this lack of consistent labeling can be unnerving: "I didn't exactly take every science class that [was on ASSIST]. And then I apply. And I think, like, 'Oh, is this gonna hurt my application [chances]?'"

As such, students suggested consistent labeling of major requirements: "There should always be reminders on what is required vs what is recommended" (P20, participant initiated).

### *Consistently Publish Annual ASSIST Reports*

ASSIST reports are published every academic year. Current ASSIST reports may or may not be different from the prior academic year's corresponding ASSIST report. For example, the 2022-2023 and 2021-2022 versions of the Orange Coast College to UC Los Angeles sociology major ASSIST report are identical in content (ASSIST, 2021b, 2022).

A common ASSIST pain point was that "some universities haven't updated [their ASSIST reports] in several years" (P22, participant initiated). When ASSIST reports aren't up to date, ASSIST "default[s]" (P17, participant initiated) to the most recently published report

---

[4] For participants who were prompted, the interview question that initiated the conversation around this software feature was along the lines of "Is it clear which ASSIST major requirements are mandatory and which are highly recommended?"

available: "sometimes [the most recently published ASSIST report will] be from like 2016. …
[So] if there's no [current report for this academic year], then you have to [use] that one [from
2016]" (P13, participant initiated). However,  ASSIST does not provide assurances or any
indication that the university is still following the default old report.

Students described feeling "confus[ed]" (P17, participant initiated) and "uneasy" (P13,
participant initiated) about possibility following "outdated information" (P22, participant
initiated) that is no longer applicable. P15 (participant initiated) elaborates: "My counselors
would always tell me, 'Oh. [ASSIST] hasn't been updated in a while. So, things might have
changed.' So that … really stressed me out. Like what if I'm missing classes now that they would
have wanted." P18 (participant initiated) provides a tangible example: "[The actual university
major requirements were] different than what ASSIST had to say. So, [my friend] spent like an
extra year for absolutely no reason because ASSIST either wasn't updated or it's missing
requirements."

P13 (participant initiated) said that universities should still publish ASSIST reports
annually even if there are no content changes in the report:

> I understand that [major requirements within ASSIST reports] don't [change] every year,
> like what courses you need to take. … It could be the same information from 2016. [If
> that is the case] they should just [publish the same report but] change the [2016] year [in
> the report] to the [current] year.

### Notification System for ASSIST Report Changes

We previously discussed the pain point of universities' ASSIST reports being out of date.
A similar but different pain point is that universities may revise *already-published* ASSIST
reports. However, there is "no indication [within the ASSIST report] that this [report] was updated" (P18, participant initiated) and "ASSIST has no way of notifying a user" (P23, participant initiated) of report revisions. For example, CSU San Marcos' ASSIST reports include this warning: "THE ARTICULATION AGREEMENT BELOW IS SUBJECT TO PERIODIC REVISION. PLEASE CONSULT A COUNSELOR EVERY SEMESTER TO OBTAIN CURRENT INFORMATION ABOUT POSSIBLE CHANGES IN THE AGREEMENT" (ASSIST, 2021a, p. 1). In other words, the contents of the 2021-2022 CSU San Marcos ASSIST reports could possibly change during the 2021-2022 academic year.

P3 (participant initiated) discusses the consequences of ASSIST's nontransparent revisions:

> [Students are unaware] that [the] class you're taking right now, it's [potentially] actually worth nothing now. … Even their counselors, who would want to tell [students], wouldn't even know. … [Advisors] don't spend all day checking every single agreement of every single university.

Due to ASSIST's nontransparent revisions, P23 (participant initiated) felt compelled to manually check ASSIST reports for revisions:

> I seen some of [the ASSIST reports] change. And so, I check[ed] it pretty frequently, like once or twice every few weeks just to make sure … I don't need a new course or … take a course that's no longer needed.

Students suggested a notification system where "if the school decides to change [the ASSIST report you are using, then] for your account to have a notification" (P5, participant initiated). P23 (participant initiated) suggested that the notification or email should also describe the revision: "a brief [description of] change[s] like '[community college] course A was changed to [community college] course B to articulate for [university] course C.'"

***Track Progress Towards Completing University Transfer Requirements***

Community colleges typically have their own internal community-college-specific software for academic planning and degree audits, which can be "awful" (P18, participant initiated) and "pretty outdated" (P22, participant initiated). Community colleges' internal academic planning software typically track students' progress for completing a community college certificate or associate degree. However, a California community college's internal software may not have the ability to track students' progress towards completing *university* major requirements; this may not be a priority for a community college due to the numerous major articulation agreements at each public California university, which are published on an annual basis and can then be revised throughout the year. For example, P24 (participant initiated) said that her community college's internal academic planning software just tracked "general transfer [requirements]" (e.g., complete 60 semester units) but it did not track the university's major requirements listed on ASSIST:

> [At] my community college, … we had our [academic] plan [software] … but that's [just] a general transfer plan. … What was required of me to fulfill to be transferable [within the community college software] versus what the [ASSIST reports from four-year] colleges actually wanted for me were different. Hence the need for ASSIST.

Similarly, ASSIST also does not track community college students' progress towards completing university major requirements: "there wasn't a way for me to verify on ASSIST like 'oh, I've already filled this requirement'" (P1, researcher initiated[5]). As such, students need to manually track their transfer progress: "I had to keep constantly going back and checking the ASSIST reports … [while] look[ing] at my [community college] class schedule and my transcripts checking 'Oh yeah. I've taken this [course listed on ASSIST]'" (P1, researcher initiated).

---

[5] For participants who were prompted, the interview question that initiated the conversation around this software feature was along the lines of "Is there any features that could be helpful for checking to make sure that you're ready to transfer based off of your completed coursework?" or "Would it be helpful or unhelpful to allow students to input the courses that they've taken into their ASSIST account?"

However, manual tracking introduces opportunities for human-error, which is compounded by ASSIST reports' ambiguity regarding how to fulfill university major requirements. P12 (researcher initiated) said "Sometimes [major requirements are] pretty straightforward [on] how to follow it. But sometimes [with other major requirements], it's kind of difficult for [students] to understand it." Likewise, P23 (participant initiated) says:

> [ASSIST does not] explicitly tells you, "Oh, you just finished your calculus requirement" for example. Sometimes it's a little unclear [if you fulfilled the requirement correctly]. … It'd be nice if the system actually says like … "you will get … calculus credit at the two UCs that you want to go to."

As touched on in the prior quote, participants suggested a software feature that would explicitly designate "which [transfer] requirements you've already fulfilled and which classes you still have to take to transfer" based on the student's planned and completed community college courses (P16, participant initiated). This feature would "save time" (P14, participant initiated) and reduce "error" (P23, participant initiated) from manual tracking of transfer progress.

**Features to Mitigate Mistakes in Term-by-Term Course Planning**

Students suggested an academic planning feature for ASSIST where they can map out term-by-term what future courses they plan to take: "you can put in which classes you take for which semester" (P2, participant initiated). P15 (participant initiated) elaborates:

You could do a customized plan, almost like a schedule. Where you can choose how many … semesters you want. … [And] even like a drag feature where you [drag and drop] all the classes [from the] ASSIST [reports or] you can [manually] type [courses] in there.

In addition, students suggested three software features that would augment the academic planning feature. This theme – features to mitigate mistakes in term-by-term course planning – has three software feature suggestions: (a) list prerequisites for community college courses, (b) list past academic terms that a community college course was offered, and (c) nudge students to prioritize completing gateway classes earlier.

### List Prerequisites for Community College Courses

ASSIST currently "doesn't show prerequisites" on their articulation agreement reports (P16, participant initiated). Consequently, some students will try to unsuccessfully register for a course without realizing that there are unmet prerequisites: "I'd realize upon registering, 'Oh, I can't take this class.' But on ASSIST, it doesn't mention anything about you need this [prerequisite]" (P9, participant initiated).

As such, research participants suggested that the "ASSIST report could be improved by … listing pre reqs of [community college] classes" (P1, participant initiated), which would help "student[s] to factor in the prerequisite classes needed to take the course" (P14, participant initiated). P5 (participant initiated) provides an example of how prerequisites can be designated in ASSIST: "there could be an asterisk or in a different color saying like, 'In order to take this class, you need to have this [prerequisite] fulfilled [or cleared].'"

### *List Past Academic Terms that a Community College Course was Offered*

ASSIST currently "doesn't tell [students] when this class is [typically] offered." (P3, participant initiated). This may cause issues because "sometimes a course may only be offered in fall semester but not in spring" (P14, participant initiated). P13 (participant initiated) describes how a mismatch between students' academic plan and the community college's actual course offerings can delay a student's time-to-transfer:

> Maybe someone's like, "Oh shoot. … I [planned to take this class] for my spring semester. But that class is [actually] only open in fall." [Consequently] they have to wait until the next semester to transfer out [because that uncompleted course is delaying them].

Another ASSIST pain point is that articulation agreement reports still list dormant courses[6] that the community college has not offered in years: "I would find it useful if courses that aren't typically offered at a [community] college aren't displayed [on ASSIST] because I don't want to choose [and plan to take] a [dormant] course that … [is actually] not offered" (P21, participant initiated).

Students suggested a feature that would allow them to "see what … semester [a course is typically offered] because I know not all classes are offered all year round" (P13, participant initiated). P16's (researcher initiated[7]) elaborates:

---

[6] As a counterargument, it may be helpful to have dormant community college courses listed on ASSIST reports on the off chance that the community college begins offering that course again.

[7] For participants who were prompted, the interview question that initiated the conversation around this software feature was along the lines of "Sometimes ASSIST will list a community college course equivalent but the community college hasn't offered that course in many years. So even though it's on the up-to-date report, the community college is no longer offering that course. So, what ASSIST functionality could address this scenario?"

> So for each course that's on the ASSIST report there could be a drop-down menu, and
> you could see which [past semesters it was offered]. For instance, like 2018 fall this was
> offered [and] or 2019 spring. And then I guess for courses that haven't been offered in
> like two or three years, there could be an asterisk or some kind of warning to the student
> that this course probably won't be offered.

In other words, historical trends of course offerings may indicate if (a) the course is active (i.e., not dormant) and (b) if there is a pattern for what academic term(s) a course is typically offered. However, historical trends are not a guarantee of future course scheduling.

### *Nudge Students to Prioritize Completing Gateway Classes Earlier*

Certain courses require many prerequisites that must be completed sequentially. We refer to this type of prerequisite sequence as gateway classes: "I call them gateway classes. Like if you don't take a certain class, [then] you're blocked from taking subsequent classes. And that just exponentially increase your time at a community college" (P23, researcher initiated[8]). For example, a university major may require community college students to complete calculus 3 to transfer. However, a student has not completed pre-calculus yet. So, the student should start their math sequence (e.g., pre-calculus, calculus 1, calculus 2, and calculus 3) as soon as possible to transfer in a timely manner. If this student delays starting their math courses, then they will be delaying their time-to-transfer by at least 4 semesters.

---

[8] For participants who were prompted, the interview question that initiated the conversation around this software feature was along the lines of "If a student is applying to transfer to many universities, there may be over 60 semester units of mandatory and recommended major requirements amongst all the universities. How could ASSIST help prioritize which courses to take?"

P23 (researcher initiated) states that some students may not know which courses to prioritize: "[Students] knew the courses they needed to take but they just got lost on when to take them. Like, what class is required for another class? Should I prioritize one or the other?"

As such, research participants suggested that ASSIST should help students prioritize gateway classes. For example, P23 (researcher initiated) said: "[ASSIST should be] prioritizing the ones that are earlier on in the [prerequisite] sequence. Maybe even having a note saying like, 'This course allows you to take course XYZ. … Prioritize taking this as soon as you're eligible.'" Likewise, P1 (participant initiated) said ASSIST should prioritize

> from an order of [introductory gateway] classes that don't really have many prerequisites [then] to [more advanced] classes that have a lot [of prerequisites]. … And then as you move on, hopefully those [earlier] classes are prerequisites for the next [classes].

To increase student understanding of gateway classes, students suggested using visualizations such as a prerequisite "flow diagram" (P23, researcher initiated) or prerequisite "timeline … Like, this class comes after this one. [And then] [t]his class comes after this one" (P1, participant initiated).

## Features to Facilitate Online Advising from Advisors and Student Peers

This theme – features to facilitate online advising from advisors and student peers – has two software feature suggestions: (a) one-on-one communication with advisors and (b) social features with student peers and advisors.

### One-on-One Communication with Advisors

Several students suggested adding features on ASSIST for "communicating [privately] with your school counselors" (P11, participant initiated). Possible communication modes included a live "chat box" (P2, participant initiated), "asynchronous" messages (P21, participant initiated), or a "Zoom [videoconference] call" (P21, participant initiated). In addition to
answering questions about "major" requirements (P5, participant initiated) or "how to use this
ASSIST stuff" (P11, participant initiated), some research participants suggested that advisors
could also check the validity of students' academic plans online instead of during in-person
meetings. For example, P9 (participant initiated) said:

> I could see each student has their own ASSIST account. And it's linked up with the
> [community] college, and the counselors have access to that … [and can] write some
> comments on … their [academic] plan … like "Oh, maybe don't take this class. Take this
> class [instead]."

### *Social Features with Student Peers and Advisors*

Several students suggested that ASSIST add social features such as discussion "forum[s]"
(P20, researcher initiated[9]), "form[ing] groups of other students that also have the same goals"
(P9, participant initiated), or "[ASSIST could have] like their own Rate My Professor where
people who've taken the course can make comments about it" (P9, participant initiated).

P20 (researcher initiated) speculated that students would be interested in social features:
"community college students are always interested in how other students transferred, especially
to big name schools." Likewise, P9 (participant initiated) said "I'd be on Discord [telling other
students] 'Oh, take this class. Don't worry about that [other class].' … I think there's definitely a
community [who would use social features on ASSIST]."

---

[9] For participants who were prompted, the interview question that initiated the conversation
around this software feature was along the lines of "What functionalities do you think would be helpful
if students had an account or profile on ASSIST?" and "Do you think that there would anything useful
about having social features with an ASSIST account?"

P9 (participant initiated) said having social features within ASSIST may resolve the limitations of current online websites: "There's websites like College Confidential and then Discord for your specific class. But there's not really a big [centralized] hub for those kind of [articulation-agreement-related social] interactions."

In addition to manually searching for forum posts in relevant subforums, P19 (researcher initiated) describes how the forum feed algorithm would display relevant forum posts on the students' "forum homepage":

> [Relevant posts] would pop up [from] people who are either trying to go to that [same] university, or [from] people who are coming from the same community college, or with the matching major. Something of those three components [community college, university, and/or major]. And then that would [lead to] more interactions from [relevant] people.

In other words, they suggested that forum posts be organized and displayed to facilitate peer-advice from similar students. P19 (researcher initiated) describes how a student would create a forum post:

> There's one [input] section where it says like "Trying to transfer to." And then another [input] section that [says] like, "[From] which community college?" and then you can start [writing the body of the] post. Like, "I'm trying to fulfill my science requirement for [the] computer science [major]. And I'm not sure if I should take the geology route or the physics route or the bio route."

Beyond student-to-student interactions, P9 (participant initiated) suggested ASSIST could also have "social media type … interactions between students and counselors." When asked directly, students said that they would find it helpful if community college advisors and transfer admissions officers[10] had verified accounts to respond to forum posts. For example, P11 (researcher initiated) said: "Yeah. That'd be really helpful. … It's an official person saying that 'Yeah. You can do that.' Or 'No. Probably not best to do that.'"

## Discussion

We identified four themes around students' software feature suggestions for ASSIST: (a) features that automate laborious academic advising tasks, (b) features to reduce ambiguity with articulation agreements, (c) features to mitigate mistakes in term-by-term course planning, and (d) features to facilitate online advising from advisors and student peers.

Some of our study participants' software feature suggestions for articulation agreement websites are novel (e.g., notification system for articulation agreement report revisions and query university major programs that the student is close to completing) but many are not novel. Similar to our study, students and academic advisors in other studies have also suggested features to reduce ambiguity/confusion with articulation agreements (Holliday-Millard, 2021; Katsinas et al., 2016; Taggart et al., 2000). Moreover, students in Katsinas et al.'s (2016) evaluation similarly suggested implementing online discussion boards to Alabama's articulation agreement website. Moreover, researchers have already begun prototyping our study participants' software feature suggestions such as academic planning prototypes (Mohamed, 2016; Shakhsi-Niaei & Abuei-Mehrizi, 2020; Xu et al., 2016) and prototypes that incorporate social features into academic advising (Bercovitz et al., 2010).

More broadly, our study participants' software feature suggestions align with prior research that found articulation agreements can be difficult to interpret (Logue et al., 2023;

---

[10] To explain, transfer admissions officers are responsible for processing applications for transfer admissions (i.e., accept or reject).

Reeping & Knight, 2021; Taylor, 2017) and are often out-of-date (Bailey et al., 2017; Schudde et al., 2020). Moreover, similar to our study participants, other research has found that students commonly enroll in multiple community colleges (Bahr, 2009, 2012); among other reasons, students do this to access transferable/articulated courses not available at their primary community college (Dunmire et al., 2011). Lastly, our results align with prior research that suggest community college academic planning can be complex (Grote et al., 2021; Lewis et al., 2016; Packard & Jeffers, 2013).

Yadamsuren et al.'s (2008) research suggests that different stakeholder groups may use community college websites in different ways and may have distinct needs. Our study participants were limited to transfer students. Accordingly, additional research is needed to gather user needs and ASSIST software feature suggestions from different stakeholder groups such as academic advisors, articulation officers, transfer admissions officers, and *current* community college students (as opposed to our study participants who were transfer students). Regarding students, user studies should consider purposefully sampling for marginalized student subgroups who have website accessibility needs (e.g., students with disabilities) or who experience transfer-related disparities (e.g., underrepresented racial/ethnic minorities).

In consultation with stakeholders, researchers should also determine the feasibility and prioritization of which new ASSIST software features to implement, if any. Researchers will also need to be mindful of *feature creep*, which refers to having too many software features that consequently make the software less usable (Page, 2009). Furthermore, researchers and administrators should consider and design guardrails against unintended consequences of suggested ASSIST software features. For example, querying university major programs that the student is close to completing may incentivize students to pursue majors or universities that they are less interested in. Another example is students may unknowingly follow erroneous peer advice from discussion forums.

Even if ASSIST did not implement any new software features, there are still non-technology-related improvements that community colleges and universities can make. For example, our study participants discussed common mistakes in academic planning, which can delay students' time-to-transfer. Accordingly, community colleges can offer workshops or online tutorials explaining how academic plans need to account for course prerequisites, gateway courses, and the course offering schedule. On the university side, our study participants complained of the ambiguity with following outdated articulation agreements. As such, universities should include explicit language on their transfer admissions websites on how community college students should proceed when the university's articulation agreements are outdated (i.e., will the university honor the most-recent-but-outdated articulation agreement?). As a last example, some study participants were paranoid that universities would change the articulation agreements' university major requirements or community college course articulations. These changes could invalidate the community college courses that students already completed. When updating degree requirements, many universities include grandfather clauses that allow *enrolled university students* to follow the old degree requirements instead of the new requirements. Similarly, when updating courses in articulation agreements, universities should consider adding grandfather clauses for *current community college students* who already completed courses under the old articulation agreement.